\documentclass{iopart}
\usepackage{epsfig}

\input{amssym.def}
\newsymbol\lesssim 132E
\newsymbol\gtrsim 1326

\newcommand{\bA}{{\bar{A}}}
\newcommand{\bC}{{\bar{C}_1}}
\newcommand{\bK}{{\bar{K}}}
\newcommand{\bQz}{{\bar{Q}_0}}
\newcommand{\bQp}{{\bar{Q}_+}}

\newcommand{\bSp}{{\bar{\Sigma}_+}}
\newcommand{\bth}{{\bar{\theta}}}
\newcommand{\be}{\begin{equation}}

\begin{document}

\title[Self-similar spherically symmetric perfect-fluid models]{
  The state space and physical interpretation of self-similar
  spherically symmetric perfect-fluid models} 
\author{
  B. J. Carr\footnote{Astronomy Unit, Queen Mary and Westfield
    College, University of London, Mile End Road, London E1 4NS,
    England. E-mail: B.J.Carr@qmw.ac.uk}, 
  A. A. Coley\footnote{Department of Mathematics and Statistics,
    Dalhousie University, Halifax, Nova Scotia, B3H 3J5, Canada. E-mail:
    aac@mscs.dal.ca}, 
  M. Goliath\footnote{Department of Physics, Stockholm
    University, Box 6730, S-113 85 Stockholm, Sweden. E-mail:
    goliath@physto.se}, 
  U. S. Nilsson\footnote{Department of Applied Mathematics,
    University of Waterloo, Waterloo, Ontario N2L 3G1, Canada. E-mail:
    unilsson@mercator.math.uwaterloo.ca} \ and
  C. Uggla\footnote{Department of Physics, University of Karlstad,
    S-651 88 Karlstad, Sweden.\\ E-mail: uggla@physto.se}}

\begin{abstract}
  The purpose of this paper is to further investigate the solution 
  space of self-similar spherically symmetric perfect-fluid models and gain
  deeper understanding of the physical aspects of these solutions. 
  We achieve this by combining the state space description of the
  homothetic approach with the use of the physically interesting
  quantities arising in the comoving approach. We focus on three types
  of models. First, we consider models that are natural inhomogeneous
  generalizations of the Friedmann Universe; such models are
  asymptotically Friedmann in their past and evolve fluctuations in
  the energy density at later times. Second, we consider so-called
  quasi-static models. This class includes models that undergo
  self-similar gravitational collapse and is important for studying
  the formation of naked singularities. If naked singularities do
  form, they have profound implications for the predictability of
  general relativity as a theory. Third, we consider a new class of
  asymptotically Minkowski self-similar spacetimes, emphasizing that
  some of them are associated with the self-similar solutions
  associated with the critical behaviour observed in recent
  gravitational collapse calculations. 
\end{abstract}

\pacs{0420, 0420J, 0440N, 9530S, 9880H}

\section{Introduction}

Spherically symmetric self-similar (SSS) perfect-fluid solutions to
Einstein's equations have attracted considerable attention over the
last two decades. Although a nearly complete classification of such
solutions has now been achieved, many of their features remain
obscure. The aim of this paper is to provide a better understanding of
the solution space and to unravel some of the physical properties of
the solutions. 

There are a number of preferred geometric structures exhibited by the
models, and these have led to several different approaches. Two of the
most commonly used are the `comoving' approach and the `homothetic'
approach. In the comoving approach, pioneered by Cahill \& Taub
\cite{art:CahillTaub1971}, the coordinates are adapted to the fluid
4-velocity vector, whereas in the homothetic approach, introduced by
Bogoyavlensky and coworkers \cite{book:Bogoyavlensky1985,art:Anile-et-al1987},
they are adapted to the homothetic vector. The main advantage of the
homothetic approach as presented by Goliath et al.
(\cite{art:Goliath-et-al1998SSS,art:Goliath-et-al1998TSS}; GNU1,
GNU2) is that the state space is compact and regular. However,
physically interesting quantities are more straightforward to obtain
in the comoving approach, as shown by Carr \& Coley
(\cite{carrdust,cc,cc2}; CC). It is therefore clear that the two
approaches are complementary, and consequently both will be used in
this paper. 

The outline of the paper is as follows. In section \ref{sec:geo} we briefly
summarize and compare the approaches of CC and GNU; the discussion here is
primarily qualitative, with most of the mathematics being relegated to
\ref{app:eqs}. One new result presented in this section is the
complete state space for the SSS models (see figure \ref{fig:match}).
In section \ref{sec:results} we focus on solutions of particular
physical interest: the asymptotically Friedmann models; the so-called
asymptotically quasi-static models, of which the naked-singularity
solutions studied by Ori \& Piran (\cite{art:OriPiran1990}; OP) are 
examples; and a class of `asymptotically Minkowski' solutions. We also
mention the `critical' solution which has been discovered in
gravitational collapse calculations, although we discuss this in more
detail elsewhere \cite{ccgnuletter}. A particularly interesting
feature of SSS solutions is the way in which their characteristics
change at certain critical equations of state and we list these
equation-of-state dependent features in section \ref{sec:eos}. Section
\ref{sec:conc} contains some concluding remarks and indicates
possible future research directions. The governing equations of both
approaches are given in \ref{app:eqs}, together with expressions
for some physically interesting quantities.

\section{Geometric structures}\label{sec:geo}

In this section, we summarize the approaches of CC and GNU and discuss
their relationship. We will consider an energy-momentum tensor of
perfect-fluid form  
\begin{equation}
  T^{\mu\nu}=\mu u^\mu u^\nu + p(g^{\mu\nu}-u^\mu u^\nu) ,
\end{equation}
where $\mu$ is the energy density, $p$ is the pressure, $u^\mu$ is
the 4-velocity of the fluid and we choose units for which $c=1$. The
only barotropic equation of state compatible with the similarity
ansatz is one of the form $p=\alpha\mu$, where $\alpha$ is a constant
\cite{art:CahillTaub1971}. Causality requires $-1\leq\alpha\leq1$. We
will confine ourselves to positive pressure and the parameter interval
$0<\alpha<1$.

\subsection{The comoving approach}\label{sec:comoving}

In the spherically symmetric situation one can introduce a time
coordinate $t$ such that surfaces of constant $t$ are orthogonal to fluid
flow lines and comoving coordinates $(r, \theta, \phi)$ which are
constant along each flow line. The metric can be written in the form
\begin{eqnarray}
  ds^{2}&=&e^{2\nu}\,dt^{2}-e^{2\lambda}\,dr^{2}-R^{2}\,d\Omega^{2} ,
  \label{lelement}\\
  d\Omega^2&=&d\theta^{2}+\sin^{2}\theta \,d\phi^{2} , \nonumber
\end{eqnarray}
where $\nu$, $\lambda$ and $R$ are functions of $r$ and $t$. The
equations have a first integral, $m(r,t)$, which can be interpreted as
the mass within comoving radius $r$ at time $t$. Unless $\alpha=0$,
this first integral 
decreases with increasing $t$ because of the work done by the
pressure. Spherically symmetric self-similar solutions can be put into a
form in which all dimensionless quantities, such as $\nu$, $\lambda$,
$S\equiv R/r$, $\mu\,t^2$ and $M\equiv m/R$, are functions only of the
dimensionless self-similar variable $z=r/t$ \cite{art:CahillTaub1971}.
Varying $z$ for a given $t$ specifies the spatial profile of various
quantities. For a given value of $r$
(i.e., for a given fluid element), it specifies their time evolution.

Some of these quantities have a particularly straightforward
physical interpretation. Thus $S$ is the scale factor, $\mu\,t^2$ specifies
the density profile at time $t$ and the mass function $M$ is related to the 
divergence of the congruence of outgoing null geodesics. Values of $z$ for
which $M=\case12$ correspond to a black hole or cosmological apparent
horizon since this divergence is zero. Another important quantity is
the function  
\begin{equation}\label{velocity}
  V(z)=e^{\lambda -\nu}z ,
\end{equation}
which represents the velocity of the spheres of constant $z$ relative
to the fluid. This should not be confused with the velocity of the
fluid with respect to a Schwarzschild foliation, the `radial
3-velocity', which we denote by $V_R$. Special significance is 
attached to values of $z$ for which $|V|=\sqrt{\alpha}$ and
$|V|=1$. The first corresponds to a sonic point, the second to a
black-hole event horizon or a cosmological particle horizon.

It is convenient to introduce a dimensionless function $x(z)$ defined by 
\begin{equation}\label{xdef}
  x(z)\equiv (4\pi \mu r^{2})^{-\alpha/(1+\alpha)}.
\end{equation}
The conservation equations $T^{\mu\nu}\!_{;\nu}=0$ can then be 
integrated to give
\begin{equation}\label{eq:eom}
  e^{\nu}=\beta x z^{2\alpha/(1+\alpha)} , \quad
  e^{-\lambda}=\gamma x^{-1/\alpha} S^{2} ,
\end{equation}
where $\beta$ and $\gamma$ are integration constants. The remaining
field equations reduce to a set of ordinary differential equations in
$x$ and $S$ (given explicitly in \ref{app:comov}). Thus, the
governing equations in the comoving approach consist of an effectively
2-dimensional system of non-autonomous ordinary differential
equations. However, it is also useful to regard the solutions as
trajectories in the 3-dimensional $(x, S, \dot{S})$ space (where a dot
denotes $z\,d/dz$). Each trajectory is parametrized by the similarity
variable $z$, so there is a 2-parameter family of spherically symmetric
self-similar solutions for a given equation-of-state parameter $\alpha$. 
Note that $z$ may be either positive or negative
but it always has the same sign as $V$. 

There is considerable arbitrariness in the selection of the
axes that specify the solution space: although CC take them to be 
$S$, $\dot{S}$ and $x$, Foglizzo \& Henriksen
(\cite{art:FoglizzoHenriksen1993}; FH) take them to be $\mu$, $V$ and a
variable linearly related to $\dot{S}$. In either case, 
the functions have an obvious physical significance. The comoving
approach thus has considerable intuitive appeal in that it affords
immediate physical insights. For example, one can write down the
metric explicitly and see immediately where the density goes infinite
and the singularities occur. 

One disadvantage of the comoving approach is that the similarity
variable $z$ can be badly behaved. For example, sometimes finite
values of $z$ correspond to zero or infinite physical distances from
the origin and sometimes zero values of $z$ correspond to non-zero
distances. Also many solutions span both negative and positive values
of $z$, which means that $z$ must jump form $+\infty$ to $-\infty$. FH
avoid this problem by using the coordinate $\xi \equiv 1/z$ but they
then have a discontinuity at $z=0$. Another disadvantage is that the
comoving approach, at least with the variables used by CC and FH,
makes rigorous proofs of general features of SSS solutions difficult. 

In $(x, S, \dot{S})$ space the sonic condition $|V|= \sqrt{\alpha}$
specifies a 2-dimensional surface. Where a curve intersects this
surface, the equations do not uniquely determine $\dot{x}$ (i.e., the
pressure gradient), so there can be a number of different solutions passing 
through the same point. However, only integral curves which pass
through a line Q on the sonic surface, the {\it sonic line}, are
`regular' in the sense that they can be extended beyond there. 
Thus the sonic surface `filters out' the small subset of solutions
that are physical. The behaviour near the sonic line can be treated as
an eigenvalue problem \cite{book:Bogoyavlensky1985,BH2}. It can be
shown that the equations permit just two values of $\dot{x}$ at each
point of Q and there will then be two corresponding values of $\dot{V}$
\cite{art:CarrYahil1990}. If the values of $\dot{V}$ are complex,
corresponding to a {\it focal} point, then the solution will still be
unphysical. If they are real, at least one of the values of $\dot{V}$
must be positive. If both values of $\dot{V}$ are positive,
corresponding to a {\it nodal} point, the smaller value is associated
with a 1-parameter family of solutions, while the larger one is
associated with an isolated solution. The eigenvector direction
associated with the 1-parameter set is referred to as {\it dominant}
or {\it primary} and the other direction as {\it secondary}. If one of
the values of $\dot{V}$ is negative, corresponding to a {\it saddle}
point, both values are associated with isolated solutions. 

\begin{figure}
  \centerline{ \hbox{\epsfig{figure=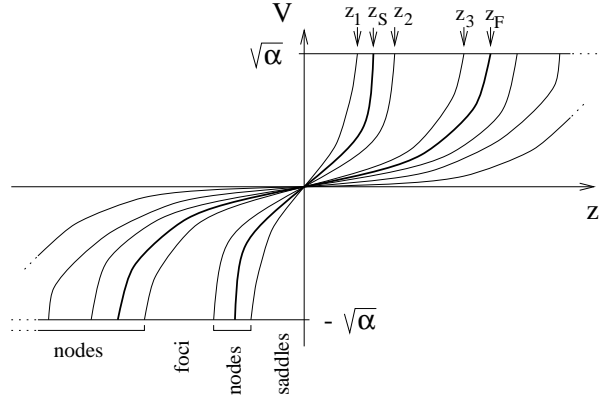, width=0.6\textwidth}}}
  \caption{The different regions of the sonic line Q in terms of a $V(z)$
    diagram. Note that there exist two sonic lines with
    $V=\pm\sqrt{\alpha}$. The curves represent the subsonic parts
    ($|V|<\sqrt{\alpha}$) of some solutions with a regular centre (to
    be discussed in more detail below). In addition, the static
    solution has been included. The thick curves correspond to
    the flat Friedmann solution and the static solution. Note that the
    orbits in one quadrant constitute the time-reverses of the orbits
    in the other quadrant. Thus, for example, the expanding Friedmann
    solution appears in the upper-right quadrant, while the
    contracting Friedmann solution is depicted in the lower-left
    quadrant.}\label{fig:Vz} 
\end{figure}

On each side of a sonic point, $\dot{x}$ may have either of the two
values.  If one chooses different values for $\dot{x}$, there will be
a discontinuity in the density, pressure and velocity gradients. If
one chooses the same value, there may still be a discontinuity in the
higher derivatives of $x$. Only the isolated solution and a single
member of the 1-parameter family of solutions  are analytic (or at
least $C^\infty$) and these are associated with the two eigenvector
directions. One can show that the part of Q for which there is a
1-parameter family of solutions corresponds to two ranges of values
for $z$ for each sign of $z$. One range $(z_1<|z|<z_2)$ lies to the
left of the Friedmann sonic point $z_{\rm F}$ and includes the static
sonic point $z_{\rm S}$; the other range $(|z|>z_3)$ includes the
Friedmann sonic point. These features are illustrated in figure
\ref{fig:Vz}. When $\alpha=\case13$, it happens that $z_2=z_{\rm S}$
and $z_3=z_{\rm F}$. 

CC have classified the $p=\alpha\mu$ SSS solutions in terms of
their asymptotic behaviour. The key steps in their analysis are: (1) a
complete analysis of the dust ($\alpha=0$) solutions, since this
provides a qualitative understanding of some of the solutions with
pressure in the supersonic ($|V|>\sqrt{\alpha}$) regime; (2) an
elucidation of the link between the $z>0$ and $z<0$ solutions; (3) a
proof that, at large and small values of $|z|$, all self-similar
solutions must have an asymptotic form in which $x$ and $S$ have a
power-law dependence on $z$; and (4) a demonstration that there are
only three {\it exact} power-law solutions -- the flat Friedmann
model, a self-similar static model and a self-similar Kantowski-Sachs
model. Prompted by results obtained by GNU, CC also find: (5) a
family of power-law solutions which are `asymptotically
Minkowski' at large $|z|$; and (6) a family of solutions which are
`asymptotically Minkowski' at a {\it finite} value of $z$ and which
have a power-law dependence on $\ln z$. These new solutions only exist
for $\alpha >1/5$; they were missed in CC's original analysis but they
were then able to extend their work to include them.

\subsection{The homothetic approach}\label{sec:hom}

When the homothetic Killing vector is non-null, the line element
adapted to the homothetic vector field can be written in a diagonal form
\cite{book:Bogoyavlensky1985}
\begin{equation}
  ds^2\!=\!\left\{\!\begin{array}{c}
    e^{2X}d\hat{s}^2\!=\!e^{2X}\!
    \left[dT^2 - D_1\!^2(T)dX^2 - D_2\!^2(T)d\Omega^2\right] \\
    e^{2T}d\bar{s}^2\!=\!e^{2T}\!
    \left[D_1\!^2(X)dT^2 - dX^2 - D_2\!^2(X)d\Omega^2\right] \\
  \end{array}
  \right. \! ,
\end{equation}
where we have distinguished between when the homothetic vector field
is spacelike ($\partial/\partial X$; upper line element) and when it
is timelike ($\partial/\partial T$; lower line element). The
unphysical spacetimes, $d{\hat s}^2$ and $d{\bar s}^2$ respectively,
are hypersurface homogeneous \cite{art:Eardley1974,art:DefriseCarter1975}
and the approach used by GNU exploits similarities with the equations
governing spatially homogeneous models
(see, e.g., \cite{book:WainwrightEllis1997}). 

By demanding that the energy density be non-negative, dominant
quantities can be determined, and bounded dimensionless variables are
obtained separately for the spatially and the timelike SSS regions.
In each of the two cases, a suitable dimensionless independent
variable is chosen, so that only one dependent variable has
dimensions. Using these variables, Einstein's equations and the
conservation equation $T^{\mu\nu}\!_{;\nu}=0$ result in a 
dynamical system in which the equation for the dimensionful variable
is decoupled. The remaining equations form 4-dimensional autonomous
dynamical systems, where the dynamical variables are related by a
constraint (see \ref{app:homoeq}). One of the variables is $V$,
defined in the comoving context above. It is related to the tilt of
the fluid flow with respect to the homothetic symmetry surfaces. 

Since GNU introduce four bounded variables, locally related by a
constraint (for similar treatments, see
\cite{art:HewittWainwright1992,art:NilssonUggla1996}), solutions are
effectively treated as orbits in a compact 3-dimensional state
space. The fact that the state space can be compactified is of great
advantage in that one can visualize all the solutions at a
glance and this will be exploited throughout this paper. However,
the physical interpretation may still be difficult. For example, parts of
the boundary of the state space are not always themselves
self-similar, as exemplified by the non-self-similar Kantowski-Sachs
state space that appears as a boundary submanifold of the spatially
SSS state space. 

\begin{table}
  \caption{Interpretation of solutions asymptotic to the given
    equilibrium points.}\label{tab:kernels}
  \begin{center}
    \begin{tabular}{@{}c@{\hspace{5mm}}l}\hline\hline
      Label & Interpretation \\ \hline
      C & Solutions with a regular centre or infinitely dispersed solutions. \\
      M, $\tilde{\rm M}$ & `Asymptotically Minkowski' solutions. \\
      K & Non-isotropic singularity solutions. \\
      F & Asymptotically Friedmann (isotropic singularity) solutions. \\
      T & Exact static solution. \\ \hline\hline
    \end{tabular}
  \end{center}
\end{table}

\begin{figure}
  \centerline{\hbox{\epsfig{figure=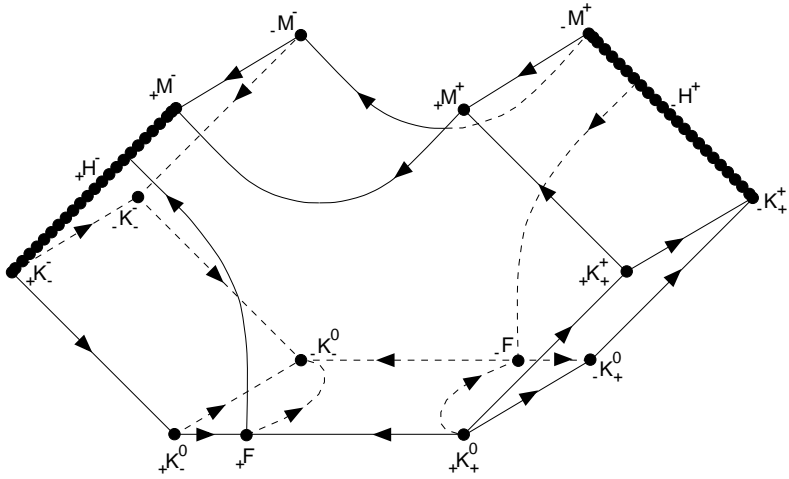, width=0.8\textwidth}}}
  \caption{The spatially SSS reduced state space for
        $\alpha\leq\case15$. For more details, see GNU1.}
  \label{fig:equiSSSless}
\end{figure}

\begin{figure}
  \centerline{ \hbox{\epsfig{figure=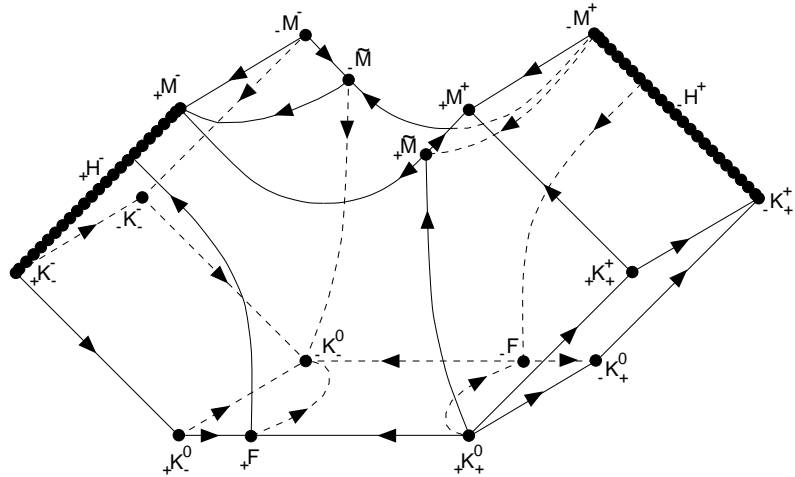, width=0.8\textwidth}}}
  \caption{The spatially SSS reduced state space for
        $\alpha>\case15$. Note that the equilibrium points
        $_\pm \tilde{\rm M}$ now have entered the spatially SSS state
        space.}\label{fig:equiSSS} 
\end{figure}

The study of the self-similar spherically symmetric solutions
now corresponds to studying the orbits in the state space of these
dynamical systems. The future and past asymptotes of the orbits are
important for understanding the corresponding models and are
associated with equilibrium points in the state space. Indeed, there
is a simple correspondence between these points and the asymptotic
classification of CC, as discussed in sections \ref{sec:comp} and
\ref{sec:results}. Such points correspond to solutions with higher
symmetry and always lie on the boundary of the state space. They often 
have a straightforward interpretation. The M points, for example,
correspond to the Minkowski solution with a certain slicing. Here we
will be interested in solutions asymptotic to the various equilibrium
points, rather than the points themselves. Thus solutions asymptotic
to the M points are `asymptotically Minkowski' and those associated
with F are asymptotic to the flat Friedmann solution. The
interpretations of different asymptotes are summarized in table
\ref{tab:kernels}. 

\begin{figure}
  \centerline{ \hbox{\epsfig{figure=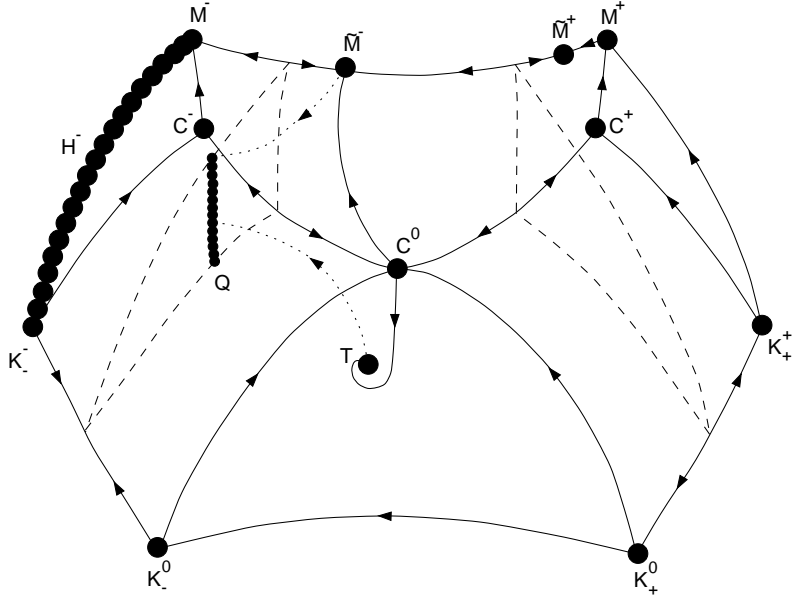, width=0.8\textwidth}}}
  \caption{The timelike SSS reduced state space for
    $\alpha<\case15$. The dashed curves indicate the boundaries of
    the sonic surfaces, located at $V=\pm\sqrt{\alpha}$. The
    bulleted line in one of these surfaces is the sonic line Q. For
    more details, see GNU2.}
  \label{fig:equiTSSless}
\end{figure}

\begin{figure}
  \centerline{ \hbox{\epsfig{figure=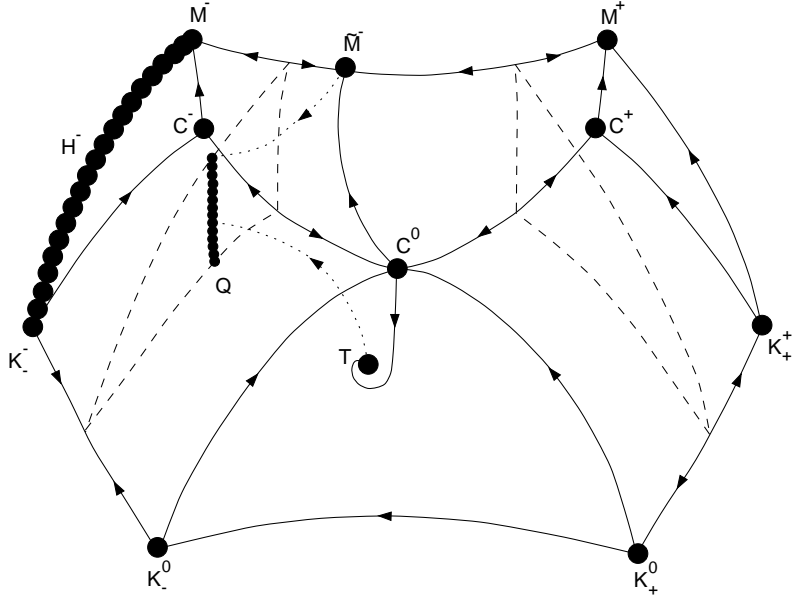, width=0.8\textwidth}}}
  \caption{The timelike SSS reduced state space for
        $\alpha\geq\case15$. Note that the equilibrium point
        $\tilde{\rm M}^+$ now has left the timelike SSS state space.}
  \label{fig:equiTSS}
\end{figure}

The labels used for the equilibrium points are sometimes accompanied by
subscripts and/or superscripts referring to the values of some of the
dynamical variables at that point. Note that models corresponding to
an equilibrium point in a spatially homogeneous slicing, such as the
flat Friedmann model, will correspond to orbits in state space in a
slicing associated with the homothety. An orbit in state space
describes a particular solution, each point on the orbit giving the
solution on a particular homothetic slice (up to a scale factor, which
is determined by the decoupled `scale' equation). 

Equilibrium points and flows along eigenvector directions for the
spatially SSS reduced state space are depicted in
figures \ref{fig:equiSSSless} and \ref{fig:equiSSS}. There is a
monotonic function for this system when $|V^{-1}|>0$ (GNU1). As there are
no invariant submanifolds with $V^{-1}=0$ in the interior, all equilibrium
points lie on the boundary of the state space. Equilibrium points and
flows along eigenvector directions for the timelike SSS reduced state
space are depicted in figures \ref{fig:equiTSSless} and
\ref{fig:equiTSS}. As for the spatially SSS region, there is a
monotonic function of the timelike SSS state space when $|V|>0$ unless
$|V|=\sqrt{\alpha}$ \cite{art:Anile-et-al1987}. 

The sonic surfaces are located at $|V|=\sqrt{\alpha}$ in the timelike
SSS regions, see figures \ref{fig:equiTSSless} and \ref{fig:equiTSS}. As
discussed in section \ref{sec:comoving}, solutions must cross the sonic
surface on the sonic line Q in order to be regular. Such a sonic line
is only present in one of the sonic surfaces. This, together with the
fact that there is a monotonic function for $|V|>0$, implies that
solutions entering the other half of the state space (the right half
of figures \ref{fig:equiTSSless} and \ref{fig:equiTSS}) are unphysical
(see GNU2). In addition, all orbits in the right half of
figure \ref{fig:equiTSSless} (figure \ref{fig:equiTSS}) pass through a
sonic surface and end at the sink $\tilde{\rm M}^+$ (${\rm M}^+$), so
that all these solutions have an irregular sonic point. It also turns
out that solutions there have negative mass. 

\begin{figure}
  \centerline{ \hbox{
      \epsfig{figure=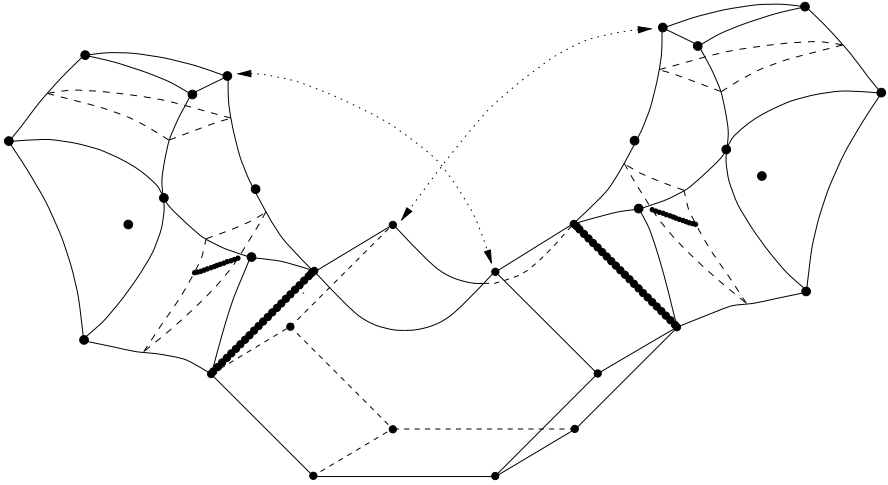, width=0.9\textwidth}}}
  \caption{Matching the spatially SSS region with two timelike SSS
    regions. Note that the equilibrium points $_\pm{\rm M}^\pm$ are
    matched with ${\rm M}^+$ in each of the timelike SSS regions, as
    indicated by the dotted arrows.}\label{fig:match} 
\end{figure}

In order to obtain a fully global picture of the solution space, the
state spaces of the spatially and the timelike SSS regions must be
matched. This is done at equilibrium points for which $|V|=1$. The
change of causality of the symmetry surfaces there corresponds to a
Cauchy horizon in these variables. In figure \ref{fig:match}, the
spatially SSS state space has been matched with two timelike SSS state
spaces along the lines of equilibrium points H. These lines of
equilibrium points are artifacts of the homothetic approach and appear
because the coordinates break down when $|V|=1$. Thus, the state space
is `pinched off' there, as seen in the figure. When matching, one must
adjust the directions of the state space flows in the regions. For
example, when following orbits from the timelike SSS region into the
spatially SSS region via $_+{\rm H}^-$, the state space flow in the
spatially SSS region will be opposite to that of figures
\ref{fig:equiSSSless} and \ref{fig:equiSSS}. One of the matchings is
done for $V=+1$. The sonic line of the corresponding timelike SSS
state space will then be located in the sonic surface at
$V=+\sqrt{\alpha}$. Consequently, due to the monotonic function
discussed above, $V<0$ is an unphysical region in this state
space. The other timelike SSS state space is attached where $V=-1$. In
that state space, the sonic line will be located at
$V=-\sqrt{\alpha}$, and $V>0$ is an unphysical region. 

The possible initial and final equilibrium points of orbits are the
following (see table \ref{tab:asympeq}): for $\alpha\leq\case15$ 
(see figures \ref{fig:equiSSSless} and \ref{fig:equiTSSless}), the only
equilibrium points that are sources or sinks are $_\pm{\rm K}_\pm^0$
in the spacelike SSS region and $\tilde{\rm M}^+$ in each
timelike SSS region. Note that the points $\tilde{\rm M}^+$ are
located in the unphysical part of each timelike SSS region. In
addition, $_\pm{\rm F}$ and ${\rm C}^0$ are separatrix-generating
equilibrium points, i.e., points for which there is a 1-parameter set
of orbits, spanning a surface in the interior of the state space. The
static solution is represented both by the equilibrium point T and by
an orbit through the interior of the timelike and spatially SSS state
spaces. This orbit is the only interior orbit connected with T. When
$\alpha>\case15$, the equilibrium points $_\pm {\rm M}^\pm$ change
stability from saddles and become sinks or sources. In addition, the
separatrix-generating points $_\pm\tilde{\rm M}$ appear in the
spatially SSS state space (see figure \ref{fig:equiSSS}). No new
features appear in the left-hand parts of figures \ref{fig:equiTSSless}
and \ref{fig:equiTSS}, i.e., the parts of the timelike SSS state space
associated with physically interesting solutions.

\begin{table}
  \caption{Possible initial and final equilibrium points of orbits in
    the matched state space, together with the number of parameters
    necessary to describe the solutions near a certain equilibrium
    point.}\label{tab:asympeq}
  \begin{center}
    \begin{tabular}{@{}lccc}\hline\hline
      & Sink/source & Separatrix-generating & One orbit\\
      Parameters & 2 & 1 & -- \\
      \hline
      $\alpha\leq\case15$ & $_\pm{\rm K}_\pm^0$, ($\tilde{\rm M}^+$) &
      $_\pm{\rm F}$, ${\rm C}^0$ & T \\
      $\alpha>\case15$ & $_\pm{\rm K}_\pm^0$, $_\pm{\rm M}^\pm$ &
      $_\pm{\rm F}$, ${\rm C}^0$, $_\pm\tilde{\rm M}$ & T \\ \hline\hline
    \end{tabular}
  \end{center}
\end{table}

\subsection{Comparison between the two approaches}\label{sec:comp}

It should be stressed that some of the differences between the two
approaches are not intrinsic but merely reflect the particular
realizations used by CC and GNU. For example, one could in principle
use compactified variables in the comoving approach and one could
also express solutions in terms of the similarity variable
in the homothetic approach. Indeed the variables used to represent the 
solutions in the two approaches must always be mathematically related
in some way (except in spacetime regions where one or both sets of
coordinates break down).  However, since the coordinate descriptions
of the spacetime are very different, physical comparison is
non-trivial.  Also, each approach suggests different `natural'
dependent and independent variables, so variables found in one
approach can be difficult to express in the other.

CC emphasize the behaviour of solutions in the asymptotic limit
$|z|\rightarrow\infty$, since the behaviour of solutions in this limit
can only take one of a few simple forms (asymptotically Friedmann,
asymptotically Kantowski-Sachs, and what they term asymptotically
`quasi-static'). However, the limit $|z|\rightarrow\infty$ does not
usually play a crucial role in the GNU analysis but just corresponds
to some region of their state space. In particular, there is no
equilibrium point corresponding to CC's asymptotically quasi-static
solutions since $|z|\rightarrow\infty$ in this case corresponds to a
surface {\it within} the state space. CC also discuss the behaviour of
solutions in the limit $z\rightarrow0$ and find that the solutions are
either exactly static or asymptotically Friedmann at the origin. 

CC place considerable emphasis on the form of the velocity function 
$V(z)$. This is useful if one wishes to identify event horizons, sonic
points and physical singularities. (In the GNU analysis these
conditions correspond respectively to the lines H, the surfaces
$|V|=\sqrt{\alpha}$ and the equilibrium points K.) However, it must be
emphasized that the $V(z)$ representation alone yields an incomplete
understanding of solutions since one is projecting them onto
a particular 2-dimensional plane, so that many physically distinct
solutions may be superposed. For example, only in the $\alpha=\case13$
(radiation) case does the region of the 3-dimensional solution space
in which the mass is negative correspond to a well-defined region of
$V(z)$ space; otherwise it depends on the third axis. CC also
emphasize the form of the scale factor $S(z)$ and the mass function
$M(z)$. These quantities provide important physical insights into the
solutions but one should bear in mind that they only represent different
2-dimensional projections of the full 3-dimensional state space.

In the homothetic approach, the variables are chosen for their mathematical
properties. This makes it possible to obtain a compact state space,
and thus a complete visual picture of the solution space. The homothetic
formulation also provides rigorous proofs and enables one to use the
insights gained from Bogoyavlensky's analysis \cite{book:Bogoyavlensky1985}.
On the other hand, this approach is rather technical, and one can only
indirectly gain physical insights. 

A complication in the diagonal homothetic approach is that the spacetime
must be covered by several coordinate patches, since the coordinates
break down when the homothetic Killing vector becomes null, $|V|=1$
(see section \ref{sec:hom}). This is physically equivalent to when the
fluid 4-velocity becomes null. Since many SSS solutions cross $|V|=1$,
one is forced to consider the relationship to other approaches if one
wants to understand the global nature of each solution. Thus, in order
to investigate when and exactly how orbits pass from one regime to the
other (e.g., timelike to spacelike) in the homothetic approach, the
dynamics are considered in the comoving approach. Going back and forth
between the homothetic and comoving approaches leads to a complete
understanding of the dynamics across the null surface and allows a
check of the internal consistency of both approaches. This is another
illustration of how the two approaches are complementary and how it is
only by using the two approaches in tandem that a full understanding
is achieved. 

Variable and coordinate transformations can be used to relate the
comoving and homothetic variables explicitly (see GNU1, GNU2
\cite{art:Goliath-et-al1998SSS,art:Goliath-et-al1998TSS}), although
care must be taken in effecting these transformations, and it should be
noted that there are some typographical errors in Appendix B of GNU1.
An alternative and perhaps a more elegant procedure is to use
expressions for dimensionless quantities, such as $\mu t^2$ and $2m/R$,
in each approach. Together with the covariant expression for $V$,
given by Cahill and Taub \cite{art:CahillTaub1971} (see equation
\ref{eq:VCT}), it is straightforward to use these to relate the
variables of the two approaches.

\subsection{The Schwarzschild approach}

We should also mention the `Schwarzschild' approach. This is useful
when matching a self-similar interior region to an asymptotically flat
(non-self-similar) exterior region. Schwarzschild coordinates are also
useful if one wishes to solve the equations of motion for the null
geodesics, as required in studying the global structure of the
solutions. Consequently it was used by OP to study naked singularities
and the cosmic censorship hypothesis in the context of SSS
solutions. However, the disadvantage of Schwarzschild coordinates is
that they involve non-physical singularities
(see \cite{book:Bogoyavlensky1985}, pp. 158--159). The transformations
between comoving and Schwarzschild coordinates are given explicitly by
OP.

\section{Solutions and their physical interpretation}\label{sec:results}

In this section, we discuss various types of solutions and their
physical properties. The strategy will be to depict a given solution
class in state space but -- in order to extract physical features --
to also plot the following physical quantities as functions of $z$: the scale
factor $S$;  the velocity function $V$; the density profile $\mu t^2$;
and the mass function $2m/R$ ($=2M$). Another quantity of physical
interest is the asymptotic energy per unit mass $E$ and this will be
useful in parametrizing solutions. The expressions for these
quantities are collected in \ref{app:physical}.

\subsection{General solution structure}\label{sec:class}

Solutions can be classified by considering their global features, as
well as their asymptotic properties. One possibility is to consider
their differentiability. Another is to use the global causal
properties of the homothetic Killing vector $\eta^\alpha$ and we now
consider this. (1) There are solutions for which $\eta^\alpha$ is
purely timelike; these develop a shock, since their orbits necessarily
end at an irregular sonic point. (2) There are solutions for which
$\eta^\alpha$ always is spacelike; these always have $|V|>1$. (3)
There are also solutions, such as the flat Friedmann solution, for which
$\eta^\alpha$ changes causality once. (4) Solutions with two causality
changes include, for example, the static solution and some
asymptotically Friedmann solutions that recollapse. (5) Solutions that
undergo three changes of the causality include those that develop a
naked singularity. For more examples, see table \ref{tab:orbiclass}.
We will now discuss some specific examples in more detail. 

\begin{table}
  \caption{Classification of orbits. T and S stand for timelike and
    spacelike, respectively.}\label{tab:orbiclass}
  \begin{center}
    \begin{tabular}{@{}ll}\hline\hline
      Causality & Example \\
      \hline
      \\
      T & Solutions with an irregular sonic point. \\
      S & \parbox[t]{0.8\linewidth}{
        Recollapsing asymptotically Friedmann solutions, always
        having $|V|>1$.\\ 
        Recollapsing asymptotically quasi-static solutions, always 
        having $|V|>1$.} \\  
      TS & \parbox[t]{0.8\linewidth}{
        Asymptotically Friedmann solutions that expand forever,
        including the flat Friedmann solution itself.\\  
        Asymptotically quasi-static solutions that collapse to form a
        black hole.} \\
      TST & The static solution. \\
      STS & \parbox[t]{0.8\linewidth}{
        Asymptotically Friedmann solutions that recollapse but have
        $|V|<1$ for part of their evolution.\\   
        Asymptotically quasi-static solutions that expand and
        recollapse, having $|V|<1$ for part of the evolution; these can
        also be of type STSTS.} \\  
      TSTS & \parbox[t]{0.8\linewidth}{
        Asymptotically quasi-static solutions that collapse to form
        naked singularities.} \\ \hline\hline 
    \end{tabular}
  \end{center}
\end{table}

\subsection{Asymptotically Friedmann solutions}\label{sec:af} 

\begin{figure}
  \centerline{\hbox{
      \epsfig{figure=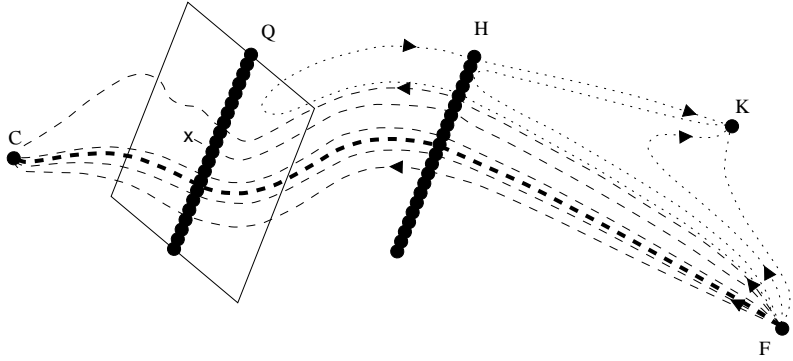, width=0.8\textwidth}}}
  \caption{Asymptotically Friedmann solutions as orbits in state
    space. Recollapsing solutions are represented by dotted
    curves. Also shown are solutions that reach the sonic surface
    (dashed curves), some of which can be continued to a dispersed
    state C. The continuations are exemplified by solutions with no
    oscillation in $V_R$ (four curves) and one oscillation in
    $V_R$ (one curve). The flat Friedmann solution is indicated by the
    thick dashed curve. `{\sf x}' marks an irregular sonic
    point.}\label{fig:afgnu} 
\end{figure}

\begin{figure}
  \centerline{ \hbox{\epsfig{figure=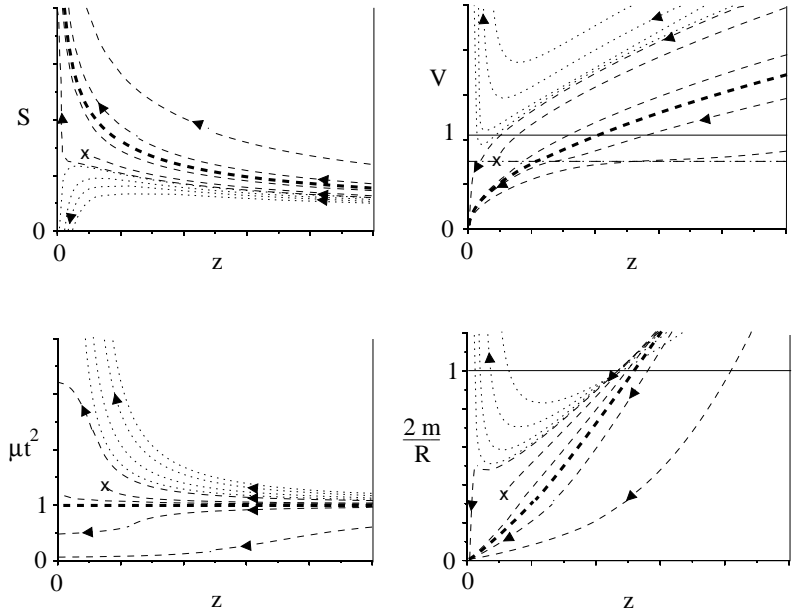, width=0.8\textwidth}}}
  \caption{Physical quantities for the asymptotically Friedmann
    solutions. The dash-dotted line in the $V(z)$ diagram
    corresponds to the sonic surface. Other designations are given in
    the caption of figure \ref{fig:afgnu}. The $\mu t^2$ diagram has
    been normalized so that the flat Friedmann solution corresponds to
    $\mu t^2=1$. The relation to the state space is as follows: the
    Friedmann point F corresponds to $z\rightarrow\infty$; the
    non-isotropic singularity (K) corresponds to $S\rightarrow0$ for
    finite values of $z$ (dotted curves only); orbits reaching $z=0$ 
    correspond to infinitely dispersed solutions (the point C; dashed
    curves only).}\label{fig:afcc} 
\end{figure}

There are two distinct 1-parameter families of asymptotically Friedmann
solutions as $|z|\rightarrow\infty$, one with $z>0$ and the other with
$z<0$. The parameter characterizing these solutions measures the
underdensity or overdensity relative to the flat Friedmann solution
and is also associated with the asymptotic energy $E$. In terms
of the GNU state space, the asymptotically Friedmann solutions start
at the F equilibrium points in the spatially SSS region (see
figure \ref{fig:afgnu}, where one of the two equivalent families of the
asymptotically Friedmann solutions is depicted). Physically
interesting quantities are illustrated in figure \ref{fig:afcc}. The
$z>0$ solutions correspond to inhomogeneous models that start from an
initial Big Bang singularity at $z=\infty$ ($t=0$) and then, as $z$
decreases, either expand to infinity or recollapse. The $z<0$
solutions are just the time-reverse of these. Depending on the value
of $E$, two qualitatively different types of solutions can be
distinguished: 

1) For solutions that are sufficiently overdense with respect to the flat
Friedmann solution (i.e., for $E$ less than some critical
negative value $E_{\rm crit}$), $|V|$ reaches a minimum and then
rises again to infinity as $|z|$ decreases, indicating the formation of
a non-isotropic singularity at which $S\rightarrow0$
for finite a value of $z$. The solutions in this class are of type S
or STS in the classification of section \ref{sec:class}. Such solutions
correspond to black holes growing at the same rate as the Universe 
\cite{art:CarrHawking1974,BH1,BH2,art:Lin-et-al1978}.
They are represented by dotted curves in figures \ref{fig:afgnu}
and \ref{fig:afcc}. Providing the minimum of $|V|$ is below 1,
there is a black-hole event horizon and a cosmological particle
horizon where $|V|=1$. Otherwise the entire Universe is inside the
black hole, although there is always an apparent horizon since CC show
that the minimum of $M$ is necessarily  below $1/2$. 

2) All solutions which are underdense or not sufficiently overdense
($E >E_{\rm crit}$) reach the sonic surface at
$|V|=\sqrt{\alpha}$. The solutions in this class are of type TS in the
classification of section \ref{sec:class}. Those which reach the sonic
line Q, and in addition have $z_1<|z|<z_2$ or $|z|>z_3$ at the sonic
point (see figure \ref{fig:Vz}), may be attached to the origin by
subsonic ($|V|<\sqrt{\alpha}$) solutions. In figures \ref{fig:afgnu}
and \ref{fig:afcc}, these asymptotically Friedmann solutions are
represented by dashed curves. Subsonic continuations that have a
regular centre have been included whenever possible. The solutions
with a regular centre are also described by a single parameter and
this is a measure of the density at the origin $z=0$. These transonic
solutions represent density fluctuations that grow at the same rate as 
the particle horizon \cite{art:CarrYahil1990}. Numerical calculations
indicate that these solutions can generally only be matched to the centre
non-analytically. While there is a continuum of regular underdense
solutions, regular overdense solutions only occur in narrow bands
(with just one solution per band being analytic). The overdense
solutions exhibit oscillations in the subsonic region, with the number
of oscillations labeling the band. Solutions with larger number of
oscillations form ever narrower bands within the one-oscillation band
(in terms of the ranges of $z$ at Q). The existence of these subsonic
bands was first pointed out by Bogoyavlensky
\cite{book:Bogoyavlensky1985} and also studied by Ori \& Piran
\cite{art:OriPiran1990}, Carr \& Yahil \cite{art:CarrYahil1990} and
GNU2. The higher bands are all nearly static near the sonic point
($z\approx z_{\rm S}$), although they deviate from the static solution
and approach the equilibrium point C as $z\rightarrow0$.

\subsection{Asymptotically quasi-static solutions}

As discussed by CC, there is exactly one self-similar static solution
for each value of $\alpha$, a 1-parameter family of solutions that
are asymptotically static (in the sense that the radial 3-velocity $V_R$
tends to zero as $|z|\rightarrow\infty$) and a 2-parameter family of
solutions that are asymptotically `quasi-static' (in the sense that
$\dot{S} \rightarrow 0$ but $V_R$ is finite). One of these
parameters can be taken to be the asymptotic energy $E$,
while the other (denoted by $D$) can be related to the value of $z$ at
the Big Bang or Big Crunch singularity. It should be emphasized that
the solutions in this class may only be close to the static solution
for part of their evolution. For example, the class includes expanding
models that finally recollapse. This is clearly illustrated by some of
the state-space orbits in figure \ref{fig:aqsgnu}. 

The key feature of these solutions is that they span both negative and
positive values of $z$. Whereas the asymptotically Friedmann solutions are
confined to $z>0$ or $z<0$ and are symmetric in $z$ (changing the
sign of $z$ just reverses the time direction), the asymptotically
quasi-static solutions are in general asymmetric in $z$ and
necessarily pass from $z=-\infty$ to $z=+\infty$. This is because the
Big Bang occurs at $z=-1/D$ (corresponding to a non-zero value of $t$)
in these solutions, so the limit $|z|\rightarrow\infty$ has no
particular physical significance. They can be interpreted as
inhomogeneous cosmological models with an advanced Big
Bang. Equivalently, for the time-reversed solutions, there is a Big
Crunch singularity at $z=+1/D$. As illustrated in
figures \ref{fig:aqsgnu} and \ref{fig:aqscc}, there are two types of
asymptotically quasi-static solutions:

\begin{figure}
  \centerline{ \hbox{
      \epsfig{figure=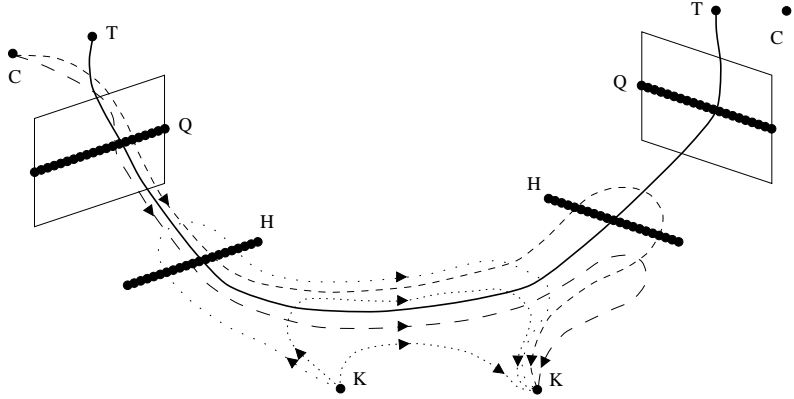, width=0.8\textwidth}}}
  \caption{Asymptotically quasi-static solutions as orbits in state
    space. Dotted curves correspond to recollapsing solutions,
    while dashed curves correspond to ever-collapsing
    solutions. In particular, the short-dashed curve corresponds to
    a solution having a naked singularity. Note that there also are
    solutions of the ever-expanding type. These go from the left-most
    K-point to the right-most C-point , in analogy with the
    ever-collapsing ones displayed here. The heavy full curve
    corresponds to the static solution. $V$ and $z$ are negative in
    the left half of the state space, and positive in the right
    half.}\label{fig:aqsgnu} 
\end{figure}

\begin{figure}
  \centerline{ \hbox{\epsfig{figure=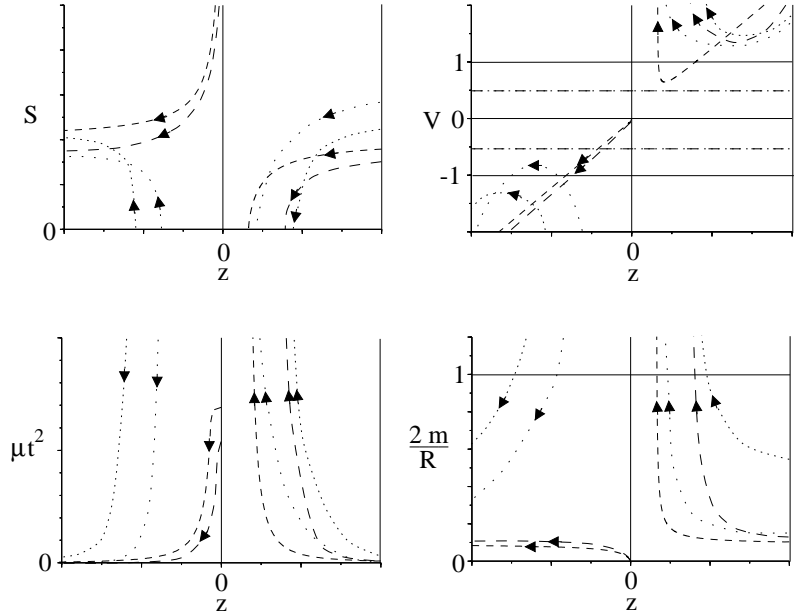, width=0.8\textwidth}}}
  \caption{Physical quantities for asymptotically
    quasi-static solutions. The dash-dotted lines in the $V(z)$
    diagram correspond to the sonic lines. Other designations are
    given in the caption of figure \ref{fig:aqsgnu}. The relation to
    the state space is as follows: $S\rightarrow0$, $\mu
    t^2\rightarrow\infty$, 
    $2m/R\rightarrow\infty$ correspond to non-isotropic singularities
    (K points); $S\rightarrow\infty$, $V\rightarrow0$, $\mu t^2$ finite,
    $2m/R\rightarrow0$ correspond to infinitely dispersed solutions
    (C points; dashed curves only). Note that $|V|\rightarrow\infty$
    both at the points K and when $z$ jumps from $-\infty$ to
    $\infty$, the latter case corresponding to a surface cutting the
    state space into two halves.}\label{fig:aqscc} 
\end{figure}

1) Expanding and recollapsing solutions. Solutions with $E$
less than some critical negative value $E_{\rm crit}(D)$ expand
from an initial singularity in the $z<0$ region and then recollapse to
another singularity in the $z>0$ region.  In terms of
the classification of section \ref{sec:class}, these are of type S, STS
or STSTS. The value of $z$ is $-1/D$ at the initial singularity but
depends on both $E$ and $D$ at the final one. These solutions
are represented by dotted curves in figures \ref{fig:aqsgnu} and
\ref{fig:aqscc}. As $z$ decreases from $-1/D$, $V$ rises from
$-\infty$, reaches a maximum below $-\sqrt{\alpha}$ but possibly above
$-1$ and then tends to the quasi-static form as $z\rightarrow -\infty$. The
solution then jumps to $z=+\infty$ and enters the $z>0$ regime. As $z$
continues to decrease, $V$ decreases to a minimum above
$\sqrt{\alpha}$ and then tends to $+\infty$ at the value of $z$
corresponding to the recollapse singularity. If the minimum of $V$ is
below $1$, one necessarily has a black-hole event horizon and a
cosmological particle horizon; this occurs if $E$ exceeds some 
negative value $E_*(D)$. The minimum of $V$ will reach
$\sqrt{\alpha}$ when $E$ reaches $E_{\rm crit}(D)$ and
this corresponds to the last recollapsing solution. One also has the
time-reverse of these solutions. 

The solutions with $E> E_{\rm crit}(D)$ are of two
types:

2a) Ever-collapsing solutions. These start out from an infinitely
dispersed state and describe the collapse of an inhomogeneous gas
cloud to a non-isotropic singularity at $z=+1/D$ (i.e., after
$t=0$). In terms of the classification of section \ref{sec:class},
these are of type TS or TSTS. The dashed curves in figures
\ref{fig:aqsgnu} and \ref{fig:aqscc} correspond to these
solutions. They start at the point C with $V=0$ at $z=0$ and then, as
$z$ decreases, reach a sonic point where $V=-\sqrt{\alpha}$. The
solution is then attached to a supersonic asymptotically quasi-static
solution at the sonic line. In this context, it should be noted that
the introduction of the `second' parameter $D$ has relatively little
effect on the form of the solutions in the subsonic regime. Indeed,
one can show that all solutions apart from the exactly static solution
must be asymptotic to the flat Friedmann solution at small $|z|$. In
particular, the models can collapse from infinity (i.e.,
$S\rightarrow\infty$ as $z\rightarrow0$ or $t\rightarrow -\infty$)
only if $E$ is positive or lies in discrete bands if negative. The
supersonic solutions pass through a Cauchy horizon (where $V=-1$)
before tending to the quasi-static form at $z=-\infty$ and jumping to
$z=+\infty$. The continued evolution for $z>0$ resembles the evolution
of case (1): as $z$ further decreases, $V$ first reaches a minimum and
then diverges to infinity when it encounters the singularity at
$z=1/D$. The minimum will be below 1 if $E$ is less than some negative
value $E_+(D)$, and in this case one necessarily has a naked
singularity, as pointed out by OP. Particular examples of this are
some of the general-relativistic Penston-Larson solutions (see figure 13
of OP). That $D$ has different values for different solutions is
clearly seen in the $S(z)$ graph in figure \ref{fig:aqscc}. 

2b) Ever-expanding solutions.
These are the time reverse of the ever-collapsing solutions.
They resemble case (1) in the $z<0$ regime but take a different
form after they have passed into the $z>0$ regime. As $z$ decreases
from $+\infty$, $V$ (rather than reaching a minimum) decreases
monotonically until it encounters a sonic point at
$V=\sqrt{\alpha}$. If this sonic point is located in the physical
ranges of the sonic line Q, the solution may be attached to the origin
$z=0$ by a subsonic solution. The behaviour in the subsonic regime is
equivalent to that discussed in case (2a). The ever-expanding
solutions are not represented in figures \ref{fig:aqsgnu} and
\ref{fig:aqscc}. However, they can be obtained from the
ever-collapsing ones (dashed curves) by letting $z\rightarrow-z$ and
$V\rightarrow-V$.

\subsection{The asymptotically Minkowski solutions for $\alpha>\case15$}

\begin{figure}
  \centerline{ \hbox{
      \epsfig{figure=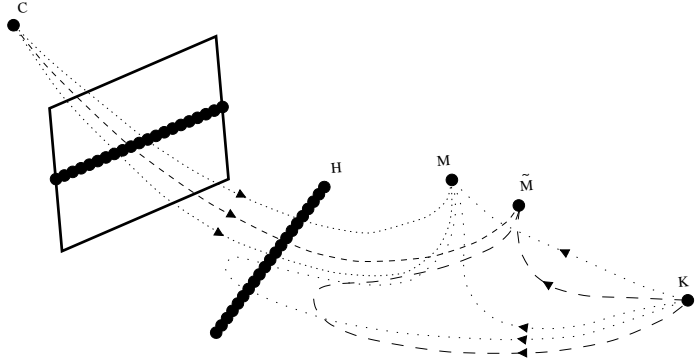, width=0.7\textwidth}}}
  \caption{Asymptotically Minkowski solutions of class A (dotted) and class
    B (dashed) as orbits in state space. Densely dotted and
    short-dashed curves correspond to regular solutions, while
    sparsely dotted and long-dashed curves correspond to singular
    solutions.}\label{fig:mgnu} 
\end{figure}

\begin{figure}
  \centerline{ \hbox{\epsfig{figure=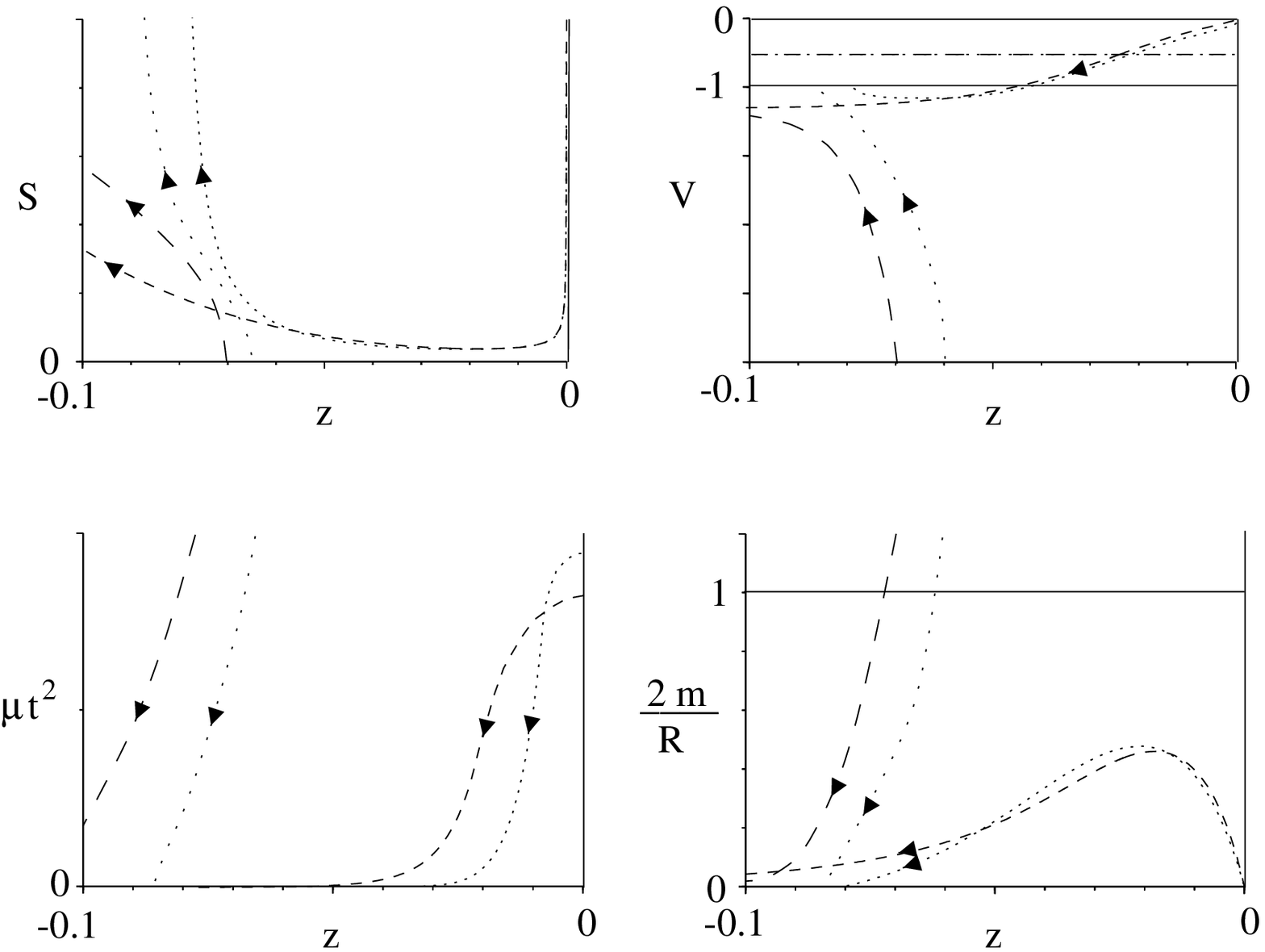, width=0.8\textwidth}}}
  \caption{Physical quantities for asymptotically Minkowski
    solutions. The dash-dotted line in the $V(z)$ diagram corresponds
    to the sonic line. Other designations are given in the caption of
    figure \ref{fig:mgnu}. The relation to the state space is as
    follows: $S\rightarrow0$, $|V|\rightarrow\infty$,
    $\mu t^2\rightarrow\infty$, $2m/R\rightarrow\infty$ corresponds to
    non-isotropic singularities (K points); $S\rightarrow\infty$,
    $V\rightarrow0$, $\mu t^2$ finite, $2m/R\rightarrow0$ as
    $z\rightarrow0$ correspond to infinitely dispersed solutions (C
    points); $S\rightarrow\infty$, $V$ finite, $\mu t^2\rightarrow0$,
    $2m/R\rightarrow0$ with $z\neq0$ correspond to the asymptotically
    Minkowski solutions (M, $\tilde{\rm M}$ points).}\label{fig:mcc} 
\end{figure}

In this section we will discuss the `asymptotically Minkowski' solutions, a
class of self-similar solutions that were only discovered very recently
and are of particular relevance to critical phenomena, see
\cite{ccgnuletter} for further details. These were missed in CC's original
asymptotic analysis, although they were able to extend their work
after GNU had demonstrated the existence of such solutions. By
combining the two approaches, we gain further physical insights into
their significance. There are actually two families of asymptotically 
Minkowski solutions and both are only physical for $\alpha>\case15$:

A) The first family is described by two parameters and is associated
with the equilibrium points M in state space. The solutions in this
family have $|V|\rightarrow1$, $S\rightarrow\infty$ and
$\mu\rightarrow0$ at some {\it finite} value $z=z_*$. In this limit, 
\begin{equation}
  \dot{V}/V=-(5\alpha-1)/(1-\alpha) .
\end{equation}
This family is illustrated by the dotted curves in figures \ref{fig:mgnu}
and \ref{fig:mcc}. Although the value of $z$ is finite at the M point, 
it should be noted out that the Schwarzschild radial distance ($R=rS$)
is infinite. This is clearly seen in figure \ref{fig:mcc}, where $S$
goes to infinity for finite values of $z$. 

B) The second family is described by one parameter and is associated
with the equilibrium points $_\pm\tilde{\rm M}$ in state space. The
solutions in this family have $|V|\rightarrow V_*>1$,
$S\rightarrow\infty$ and $\mu\rightarrow0$ as $z\rightarrow\infty$.
The expression for $V_*$ is
\begin{equation}
  V_*=\frac{\alpha(\alpha+1)+\sqrt{\alpha(\alpha^3-\alpha^2+3\alpha+1)}}
  {1-\alpha} .
\end{equation}
In figures \ref{fig:mgnu} and \ref{fig:mcc}, the dashed curves
correspond to this family of solutions. 

Both families of solutions have $2m/R\rightarrow0$ as a solution approaches
M or $\tilde{\rm M}$. However, even though $2m/R\rightarrow0$ as one
approaches M, the mass $m$ need not vanish. It is important to realize
that these solutions are not asymptotically flat in the usual
sense. Rather, they are perfect-fluid spacetimes for which the
Minkowski geometry is obtained asymptotically along certain coordinate
lines. This is reminiscent of the open Friedmann solutions, which
asymptotically approaches the Milne model along certain timelines
(see, e.g., \cite{book:WainwrightEllis1997}).

As can be seen in figures \ref{fig:mgnu} and \ref{fig:mcc}, both
families contain solutions that can be connected either to the origin
$z=0$ (via a sonic point) or to a non-isotropic singularity (for which
$S\rightarrow0$ at finite values $z_{\rm K}$). The former are of type TS,
while the latter are of type S or STS, using the classification of
section \ref{sec:class}. For both families, the limit
$S\rightarrow\infty$ can be regarded as corresponding to an infinitely
dispersed state, analogous to the late stage of an open Friedmann
model, which is described by the Milne solution, see, e.g., 
\cite{book:WainwrightEllis1997}. 

Note that there is a 1-parameter family of non-isotropic singularity
solutions for each value of  $z_{\rm K}$ (i.e., one can take one of the two
parameters that specify the solutions near K to be the value of the
similarity variable at K). Most of these will be asymptotic
to either a type A asymptotically Minkowski solution or a quasi-static
solution. However, for sufficiently large values of $|z_{\rm K}|$,
there will also be one type B asymptotically Minkowski solution and one
asymptotically Friedmann solution, all connected with K.
The limiting value of $z_{\rm K}$ is the same in each case, reflecting
the fact that the conditions at large values of $|z|$ have little
influence on what happens near the singularity. This limiting value
corresponds to the splitting of the asymptotically Friedmann and
asymptotically $\tilde{\rm M}$ submanifolds: the orbits belonging to
each separatrix surface can be described by one parameter. For some
ranges of this parameter they are asymptotic to a K 
equilibrium point, corresponding to a non-isotropic singularity; for
other ranges they have a sonic point. This is illustrated for the
asymptotically Minkowski solutions in figure \ref{fig:mgnu}, where
dashed curves correspond to type B solutions. The analogous behaviour
of the asymptotically Friedmann solutions can be seen in figure
\ref{fig:afgnu}. 

\newpage

\section{Examples of equation-of-state dependent features}\label{sec:eos}

There are several features of the solutions that depend on the
equation-of-state parameter $\alpha$. In particular, the behaviour at
the sonic line is greatly affected by the equation of state. Below, we
list some of these equation-of-state dependent features.  

\begin{itemize}
  \item $\alpha\approx0.04$: The boundary of the underdense
  subsonic band (see subsection \ref{sec:af}) changes with increasing
  $\alpha$ from being the general-relativistic Penston-Larson
  solution to the degenerate node in the nodal region containing the
  flat Friedmann solution (i.e., $z=z_3$).
  \item $\alpha\approx0.11$: The upper boundary (as defined by
  OP) of the first overdense subsonic band changes with increasing
  $\alpha$ from being a secondary eigenvector direction in the nodal region
  containing the static solution to the degenerate node of that
  nodal region (i.e., $z=z_2$).
  \item $\alpha=\case15$: The equilibrium points $\tilde{\rm M}^+$
  leave each timelike SSS state space and enter the spatially SSS 
  state space as the equilibrium points $_\pm\tilde{\rm M}$.
  \item $\alpha\approx0.28$: The critical solution, which
  corresponds to the lower boundary of the first overdense subsonic
  band, changes character with increasing $\alpha$ from a solution
  with a (generally irregular) second sonic point to a solution
  whose corresponding orbit ends at an M point in the spatially SSS
  state space. For the threshold value of $\alpha$, the critical
  solution ends at an $\tilde{\rm M}$ point.
  \item $\alpha=\case13$: The degenerate nodes of the nodal
  regions of the sonic line coincide with, respectively, the flat
  Friedmann solution and the static solution (i.e., $z_{\rm F}=z_3$ and
  $z_{\rm S}=z_2$).\\ 
  The eigenvector direction at the sonic line of the
  flat Friedmann solution changes with increasing $\alpha$ from being a
  dominant direction to a secondary direction. For the static
  solution, the situation is reversed: the eigenvector direction
  going from being secondary to dominant. \\
  The boundary of the underdense subsonic band changes
  with increasing $\alpha$ from being the degenerate node in the
  nodal region containing the flat Friedmann solution to the
  flat Friedmann solution itself (i.e., $z=z_{\rm F}$).
  \item $\alpha\approx0.41$: The eigenvector direction associated
  with the critical solution at the sonic line changes with
  increasing $\alpha$ from being an attractive eigenvector direction
  in the saddle region of the sonic line to a secondary eigenvector
  direction in the nodal region containing the static solution
  (i.e., $z_1<z<z_2$).
  \item $\alpha\approx0.45$: The overdense subsonic bands
  overlap, so that the band structure degenerates to only
  two bands: the underdense band and one overdense band.
  \item $\alpha\approx0.61$: The zero in the radial 3-velocity
  $V_R$ of the critical solution changes with increasing $\alpha$
  from being subsonic to being supersonic (i.e., for this value of
  $\alpha$, $V_R=0$ when $|V|=\sqrt{\alpha}$).
  \item $\alpha\approx0.89$: The eigenvector direction associated
  with the critical solution at the sonic line coincides with the
  degenerate node in the nodal region associated with the static
  solution (i.e., $z=z_2$), and changes with increasing $\alpha$
  from being a secondary eigenvector direction to a dominant
  eigenvector direction.
\end{itemize}

\section{Concluding remarks}\label{sec:conc}

It is clear from the present work that we are close to a complete 
understanding of the self-similar spherically symmetric solutions.
This paper has emphasized the advantages of a combined comoving and
homothetic approach. This is par\-ticularly clear in the
investigations of the asymptotically quasi-static solutions and
asymptotically Minkowski solutions.

An important application is to study the critical solution of
spherically symmetric perfect-fluid collapse. This has been done in a
separate paper \cite{ccgnuletter}, where it is shown that the critical
solution is the unique self-similar solution which is analytic at the
sonic point, has a regular centre, and contains one collapsing region
surrounded by a dispersing exterior. As indicated in section
\ref{sec:eos}, this solution is a member of the asymptotically
quasi-static class for $0<\alpha\lesssim0.28$, a member of the
asymptotically Minkowski class of type B for $\alpha\approx0.28$, and a
member of the asymptotically Minkowski class of type A for
$0.28\lesssim\alpha<1$.

Possible future investigations involve studying these models in more
general contexts. One might ask if various features discussed here are
structurally stable, i.e., how do other geometries and matter sources
affect the behaviour? For example, one might further examine the role
of the self-similar solutions in the complete space of spherically
symmetric solutions. It also remains to consider the stiff fluid
($\alpha=1$) case, which is of relevance in the study of massless
scalar fields. There are numerous bifurcations associated with the
limit $\alpha\rightarrow1$ and this explains the numerical problems
one encounters in this limit. However, we do not see any {\it a priori}
analytic difficulties in studying these models using the methods
presented in this paper.

\section*{Acknowledgements}

We would like to thank Dave Neilsen for helpful comments. AAC was
supported by the Natural Sciences and Engineering Research Council of
Canada. CU was supported by the Swedish Natural Research Council.

\appendix

\section{Equations}\label{app:eqs}

This appendix collects the basic equations in each approach, together
with expressions for the physically interesting quantities that we
have used.

\subsection{Comoving equations}\label{app:comov}

The field equations (with $G=c=1$) reduce to a set of ordinary
differential equations in $x$ and $S$: 
\begin{equation}\label{eq:ddotS}
  \ddot{S}+\dot{S}+\left(\frac{2}{1+\alpha}\frac{\dot{S}}{S}
  -\frac{1}{\alpha}\frac{\dot{x}}{x}\right)
  [S+(1+\alpha)\dot{S}]=0,
\end{equation}
\begin{eqnarray}
  & &\left(\frac{2\alpha\gamma^{2}}{1+\alpha}\right)S^{4}
  +\frac{2}{\beta^{2}}\frac{\dot{S}}{S}\,x^{2(1-\alpha)/\alpha}
  z^{2(1-\alpha)/(1+\alpha)} - \nonumber \\
  & & \gamma^{2}S^{4}\,\frac{\dot{x}}{x}
  \left(\frac{V^{2}}{\alpha}-1\right) = 
  (1+\alpha)x^{(1-\alpha)/\alpha},\label{eq:dotx}
\end{eqnarray}
\begin{equation}\label{eq:M1}
  M=S^{2}x^{-(1+\alpha)/\alpha}\left[1+(1+\alpha)\frac{\dot{S}}{S}\right],
\end{equation}
\begin{eqnarray}
  M&=&\frac{1}{2}+\frac{1}{2\beta^{2}}x^{-2}z^{2(1-\alpha)/(1+\alpha)}
  \dot{S}^{2}- \nonumber \\
  & &\frac{1}{2}\gamma^{2}x^{-(2/\alpha)}S^{6}
  \left(1+\frac{\dot{S}}{S}\right)^{2}, \label{eq:M2}
\end{eqnarray}
where an overdot denotes $z\,d/dz$.
At any point in the $(x,S,\dot{S})$ space, for a fixed value of
$\alpha$, equations (\ref{eq:M1}) and (\ref{eq:M2}) provide a constraint
that gives the value of $z$; equation (\ref{eq:dotx}) then gives the value
of $\dot{x}$ unless $|V|=\sqrt{\alpha}$ and equation (\ref{eq:ddotS}) gives
the value of $\ddot{S}$. Thus the equations generate a vector field
$(\dot{x},\dot{S},\ddot{S})$ and this specifies an integral curve at
each point of the 3-dimensional space, each curve representing one
particular similarity solution. 

Where $|V|=\sqrt{\alpha}$, equations (\ref{eq:M1}, \ref{eq:M2}) together
with the expression for $V(z)$ [see equation (\ref{eq:V})] allow one to
express $\dot{S}$ in terms of $x$ and $S$, corresponding to a surface
in $(x,S,\dot{S})$ space. Integral curves intersect
$|V|=\sqrt{\alpha}$ in a physically reasonable manner only if 
\begin{eqnarray}
  & &\left(\frac{2\alpha\gamma^{2}}{1+\alpha}\right)S^{4}
  +\frac{2}{\beta^{2}}\frac{\dot{S}}{S}\,x^{(2-2\alpha)/\alpha}
  z^{(2-2\alpha)/(1+\alpha)}  \nonumber \\
  &=& (1+\alpha)x^{(1-\alpha)/\alpha},\label{eq:Q}
\end{eqnarray}
since otherwise the value of $\dot{x}$ and hence the pressure, density
and velocity gradient diverge there. Since equation (\ref{eq:Q})
corresponds to another 2-dimensional surface in $(x, S, \dot{S})$
space, this will intersect the surface $|V|=\sqrt{\alpha}$ on a line
Q, the sonic line.

\subsection{Homothetic equations}\label{app:homoeq}

Here, we briefly summarize the quantities and equations used in the
diagonal homothetic approach. In the spatially self-similar case, the
dependent variables ($\bQz,\bQp,\bC,v$) are defined in terms of the
line element 
\begin{equation}
  ds^2=e^{2X}\,\frac{3}{Y^2}
  \left[d\tau^2-\frac{dX^2}{\bC^2}-\frac{d\Omega^2}{(1-\bQz\!^2)}\right] ,
\end{equation}
together with $\bQp=-D_1^\prime/D_1=Y^\prime/Y+\bC^\prime/\bC$ and
$v=1/V$, where ${}^\prime=d/d\tau$. $Y$ is given by the decoupled equation
\begin{equation}
  Y^\prime =-\left\{\bQp + \bQz
  \left[2\bQp^2 + \frac{1+\alpha}{1+\alpha v^2}\Omega_n\right]\right\} Y .
\end{equation}
The Friedmann equation defines $\Omega_n$:
\begin{equation}
  \Omega_n=1-\bQp^2-\bC^2 .
\end{equation}
The constraint takes the form
\begin{equation}
  G = (1+\alpha) v \Omega_n - 2\left[1+\alpha v^2\right]\bQp\bC = 0 .
\end{equation}
The reduced set of evolution equations is
\begin{eqnarray}
  \bQz\!^\prime &=&
  -(1-\bQz\!^2) \left[\bQp\!^2-\bC\!^2 +
  \frac{\alpha (1-v^2)}{1+\alpha v^2} \Omega_n \right] ,\nonumber\\
  \bQp\!^\prime &=&
  -\bQz\bQp\left[2(1-\bQp\!^2)-\frac{(1+\alpha)\Omega_n}{1+\alpha v^2}\right]
  \nonumber\\
  &-&\frac{1}{2}\frac{(1-\alpha)+(3\alpha+1)v^2}{1+\alpha v^2} \Omega_n ,
  \nonumber \\
  \bC\!^\prime &=&
  2\bC\left[\bQp + \bQz\bQp\!^2 +
  \frac{1}{2}\frac{1+\alpha}{1+\alpha v^2}\bQz\Omega_n \right] ,\nonumber\\
  v^\prime &=&
  \frac{1-v^2}{(1+\alpha)\left[1 - \alpha v^2\right]} 
  \left\{(1+\alpha)\left[ 2\alpha\bQz + (1+\alpha)\bQp \right] v \right. 
  \nonumber \\
  &+&\left.\left[\alpha (3\alpha+1)v^2 - (1-\alpha)\right] \bC\right\} .
\end{eqnarray}

In the timelike self-similar case, the dependent variables ($\bSp,\bA,\bK,V$)
are defined in terms of the line element 
\begin{equation}
  ds^2=e^{2T}\,\frac{3}{\bth^2}
  \left[\frac{dT^2}{\bA^2}-d\xi^2-\frac{d\Omega^2}{\bK}\right] ,
\end{equation}
together with $\bSp=-D_1^\prime/D_1=\bth^\prime/\bth+\bA^\prime/\bA$,  
where ${}^\prime=d/d\xi$. $\bth$ is given by the decoupled equation 
\begin{equation}
  \bth^\prime = -\left[1 + \bSp(1+\bSp) - \bA^2 - 
  \frac{\alpha (1-V^2)}{1+\alpha V^2}\Omega_t\right] \bth .
\end{equation}
The Friedmann equation defines $\Omega_t$:
\begin{equation}
  \Omega_t=\frac{1+\alpha V^2}{V^2+\alpha}\left(1-\bSp^2-\bA^2-\bK\right) .
\end{equation}
The constraint takes the form
\begin{equation}
    G = (1+\alpha)V\Omega_t - 2\left[1+\alpha V^2\right]\bSp \bA = 0 .
\end{equation}
The reduced set of evolution equations is
\begin{eqnarray}
  \bSp\!^\prime &=&
  -\bSp \left[1-\bSp\!^2+\bA^2 +
  \frac{\alpha (1-V^2)}{1+\alpha V^2}\Omega_t\right] \nonumber\\
  &-&\frac{1}{2}\frac{(3\alpha+1) + (1-\alpha)V^2}{1+\alpha V^2}\Omega_t
  , \label{eq:dbSp} \nonumber \\
  \bA^\prime &=&
  \left[1+2\bSp+\bSp\!^2-\bA^2 -
  \frac{\alpha (1-V^2)}{1+\alpha V^2}\Omega_t\right]\bA , \nonumber\\
  \bK^\prime &=&
  2\left[\bSp^2-\bA^2 -
  \frac{\alpha (1-V^2)}{1+\alpha V^2}\Omega_t\right] \bK , \nonumber\\
  V^\prime &=&
  \frac{1-V^2}{(1+\alpha)[V^2-\alpha]} 
  \left\{(1+\alpha)\left[2\alpha + (1+\alpha)\bSp\right] V 
  \right. \nonumber \\
  &+& \left. \left[\alpha (3\alpha+1) - (1-\alpha)V^2\right] \bA\right\} .
\end{eqnarray}
Note that $V$ was denoted by $u$ in GNU2.

\subsection{Physical quantities}\label{app:physical} 

Here, we give the expressions for the different physically interesting
quantities we have used.
First we note that in the homothetic approach, the similarity variable
$z=r/t$ is 
\begin{eqnarray}
  z&\propto&{\rm sgn }(v)e^{-\sqrt{3}\int d\tau Y^{-1}}\nonumber\\
  &\propto&{\rm sgn }(V)e^{\sqrt{3}\int d\xi \bth^{-1}},
\end{eqnarray}
where equation (\ref{velocity}) and the transformations between the
comoving and 
homothetic approaches (see GNU1,2) have been used.

The scale factor $S\equiv R/r$ is one of CC's dependent variables. In
the homothetic approach, it is determined by
\begin{eqnarray}
  S^2(z)&\propto&|z|^{-(1-\alpha)/(1+\alpha)}\frac{\bC}{|\bQp|}\nonumber\\
  &\propto&|z|^{-(1-\alpha)/(1+\alpha)}\frac{\bA}{|\bSp|} .
\end{eqnarray}

GNU use the velocity function (or its inverse) as one of their
dependent variables. It can be covariantly expressed in terms of the
fluid 4-velocity $u^\alpha$ and the normal to the homothetic symmetry
surfaces $n^\alpha$ as follows \cite{art:CahillTaub1971}
\begin{equation}\label{eq:VCT}
  V=\frac{u^\alpha n_\alpha}{\sqrt{(u^\alpha n_\alpha)^2 -
      n^\alpha n_\alpha}} .
\end{equation}
In the comoving variables used here, $V$ takes the form
\begin{equation}\label{eq:V}
  V=(\beta\gamma)^{-1}x^{(1-\alpha)/\alpha}S^{-2}
  z^{(1-\alpha)/(1+\alpha)} .
\end{equation}

The expressions for the density profile are
\begin{eqnarray}
  \mu\,t^2&=& \frac{1}{4\pi}z^{-2}x^{-(1+\alpha)/\alpha} \nonumber\\
  &=& S^{-2}z^{-2}\frac{1-v^2}{1+\alpha v^2}\frac{\Omega_n}{1-\bQz^2}
  \nonumber\\
  &=& S^{-2}z^{-2}\frac{1-V^2}{1+\alpha V^2}\frac{\Omega_t}{\bK} .
\end{eqnarray}

The mass function is defined as $M\equiv m/R$ where 
\begin{equation}\label{firstint}
  m(r,t)=\mbox{$\frac{1}{2}$} R\left[ 1+e^{-2\nu}
  \left(\frac{\partial R}{\partial t}
  \right)^{2} - e^{-2\lambda}\left(\frac{\partial R}
  {\partial r}\right)^{2}\right] ,
\end{equation}
or equivalently
\begin{equation}\label{massfunct}
  m(r,t)=4\pi\int_{0}^{r}\mu R^{2}\frac{\partial R}{\partial r'}\,dr' .
\end{equation}
In the comoving approach, it is given by either of equations
(\ref{eq:M1},\ref{eq:M2}), while in the homothetic approach it takes
the forms
\begin{eqnarray}
  M&=&\frac{1+2\bQz\bQp+\bQp^2-\bC^2}{2(1-\bQz^2)} \nonumber\\
  &=&\frac{\bK-(1+\bSp)^2+\bA^2}{2\bK} .
\end{eqnarray}

Finally, the asymptotic energy per unit mass $E$ is the limit
$|z|\rightarrow\infty$ of the dimensionless quantity ${\cal E}$, which
represents the total energy per unit mass of a comoving shell:
\begin{eqnarray}
  {\cal E}&=&\frac{1}{2}\left\{
  \gamma^2x^{-2/\alpha}S^6\left(1+\frac{\dot{S}}{S}\right)-1
  \right\} \nonumber\\
  &=&\frac{1}{2}\left\{
  \frac{[v(\bQz+\bQp)+\bC]^2}{(1-\bQz^2)(1-v^2)}-1
  \right\} \nonumber\\
  &=&\frac{1}{2}\left\{
  \frac{(1+\bSp+\bA V)^2}{\bK(1-V^2)}-1
  \right\} .
\end{eqnarray}

\end{document}